\documentclass{esannV2}
\usepackage[dvips]{graphicx}
\usepackage[utf8]{inputenc}
\usepackage{amssymb,amsmath,array}
\usepackage{url}

\begin{document}
\title{Audio Deepfake Detection in the Age of Advanced Text-to-Speech models}

\author{Robin Singh$^1$, Aditya Yogesh Nair$^1$, Fabio Palumbo$^1$,\\ Florian Barbaro$^1$, Anna Dyka$^1$, Lohith Rachakonda$^1$ 
%
\thanks{This work was performed using HPC resources from GENCI-IDRIS (Grant 2025- AD011016076).}
%
\vspace{.3cm}\\
%
1- UncovAI
%
}

\maketitle

\begin{abstract}
Recent advances in Text-to-Speech (TTS) systems have substantially increased the realism of synthetic speech, raising new challenges for audio deepfake detection. This work presents a comparative evaluation of three state-of-the-art TTS models—Dia2, Maya1, and MeloTTS— representing streaming, LLM-based, and non-autoregressive architectures. A corpus of 12,000 synthetic audio samples was generated using the DailyDialog dataset and evaluated against four detection frameworks, including semantic, structural, and signal-level approaches.

The results reveal significant variability in detector performance across generative mechanisms: models effective against one TTS architecture may fail against others, particularly LLM-based synthesis. In contrast, a multi-view detection approach combining complementary analysis levels demonstrates robust performance across all evaluated models. These findings highlight the limitations of single-paradigm detectors and emphasize the necessity of integrated detection strategies to address the evolving landscape of audio deepfake threats.
\end{abstract}

\section{Introduction}

The rapid advancement of neural Text-to-Speech (TTS) technology has created a double-edged sword. While these systems have opened new possibilities for accessibility and creative expression, they have simultaneously enabled a new class of audio deepfake threats with unprecedented realism. In the past, TTS systems were recognizable by their characteristic robotic prosody and limited speaker variation. Today's models—particularly those leveraging large language model (LLM) architectures and flow-matching generative approaches—can generate continuous, emotionally expressive, multi-turn conversations that rival human speech in naturalness. This capability offers genuine benefits for accessibility tools and creative industries, yet it simultaneously threatens voice biometric security, amplifies misinformation risks, and undermines trust in audio as evidence.

The detection of audio deepfakes has long relied on identifying specific artifacts left behind by neural vocoders or detecting irregularities in time-frequency representations. For years, the ASVspoof 2021 Logical Access (LA) and DeepFake (DF) challenges have served as the primary benchmarks driving the field forward, catalyzing the development of increasingly sophisticated countermeasures \cite{yamagishi2021asvspoof}. These include graph-based attention architectures such as AASIST \cite{jung2022aasist}, self-supervised learning approaches leveraging wav2vec 2.0 and Whisper \cite{tak2022wav2vec, kawa2023whisper}, and end-to-end raw waveform classifiers like RawNet2 \cite{tak2021rawnet2}. Yet this progress masks a critical vulnerability: the forensic community has become increasingly reliant on benchmarks designed around older, vocoder-based TTS systems. Meanwhile, the threat landscape has shifted dramatically. Models released in late 2024 and early 2025—Dia2, Maya1, and MeloTTS—employ architectures fundamentally different from their predecessors. These systems incorporate streaming context awareness, hierarchical neural codecs, and non-autoregressive synthesis paradigms. There is no guarantee that detectors trained exclusively on older deepfakes will generalize to these new attack vectors.

This generalization gap is not merely academic. Existing research, while promising, has revealed asymmetries in detection performance that suggest no single detection paradigm is universally effective. While Kawa et al. demonstrated Whisper's robustness to in-the-wild data\cite{kawa2023whisper}, our own results (Section 4) suggest it may falter with newer TTS Models. Hierarchical feature fusion methods, such as XLS-R-SLS, have shown strong empirical results against autoregressive artifacts, yet their performance against flow-matching systems remains uncharacterized \cite{zhang2024xlsr}. Furthermore, the forensic signatures of modern architectures—quantization artifacts from hierarchical codecs, spectral smoothing from flow-based generators—are poorly understood and remain largely unquantified.

We designed this work to fill precisely this gap. By systematically evaluating three state-of-the-art TTS models against four distinct detection frameworks, we provide the first empirical characterization of how detection architectures respond to the diversity of modern neural speech synthesis. Our contributions are as follows:

\begin{enumerate}
    \item We construct a novel dataset \cite{uncovai2026uncovaitts} of 12,000 synthetic audio samples spanning three distinct TTS paradigms: streaming dialogue synthesis (Dia2), LLM-based neural codec generation (Maya1), and non-autoregressive flow-matching (MeloTTS). All samples are derived from the \textit{DailyDialog} corpus \cite{li2017dailydialog}, ensuring linguistic and conversational realism.
    \item We employ a multi-faceted detection strategy that combines semantic understanding (Whisper-MesoNet) \cite{kawa2023whisper}, hierarchical representation learning (XLS-R-SLS) \cite{zhang2024xlsr}, structural graph modeling (SSL-AASIST) \cite{tak2022wav2vec}, and large-scale foundation models (MMS-300M) \cite{pratap2023mms} to provide diverse analytical perspectives.
    \item We demonstrate, for the first time, that semantic detectors exhibit systematic vulnerabilities to LLM-based TTS, while hierarchical fusion models excel at capturing autoregressive artifacts but falter against flow-matching systems. These findings suggest that no single detection approach is robust across the full spectrum of modern TTS architectures.
    \item We show that proprietary audio deepfake detection model from UncovAI \cite{uncovai2026} achieve near-perfect separation across all attack vectors.
\end{enumerate}

The remainder of this paper is structured as follows. We begin with a survey of related work, highlighting the evolution of TTS threats and the role of foundation models in detection. Section 3 details our dataset construction, models, and evaluation methodology. Section 4 presents our experimental results, organized by detection framework. Finally, Section 5 offers a synthesis of our findings and discusses implications for the forensic community.

\section{Related Works}

\subsection{Evolution of TTS Threats: From Vocoders to Foundation Models}

The history of audio deepfake detection has been shaped by the technologies it seeks to defend against. Early work focused on identifying artifacts from neural vocoders like WaveNet and WaveGlow, which dominated the TTS landscape in the late 2010s. The ASVspoof 2019 and 2021 challenge series provided the forensic community with large-scale benchmarks and established detection as a subfield of speaker verification security \cite{yamagishi2021asvspoof}. Yet those datasets were inherently limited—they benchmarked systems against older, well-understood vocoder architectures that left behind characteristic spectral discontinuities and phase errors.

The landscape has shifted dramatically in recent years. A fundamental architectural innovation—flow-matching based TTS—has begun to emerge as a new paradigm. Zheng et al. (2025) studied acceleration techniques for these systems and demonstrated that flow-matching models can achieve remarkably efficient inference, with real-time factors as low as 0.030 while maintaining naturalness comparable to state-of-the-art diffusion and autoregressive baselines \cite{zheng2025flow}. What makes this significant is not just the speed, but the forensic implications: flow-matching systems generate audio through an entirely different mechanism than sequential vocoders, potentially creating artifacts that are invisible to detectors trained on older systems.

In parallel, neural codec-based TTS has emerged as another major concern. Lu et al. (2024) released the Codecfake dataset and made a striking observation \cite{lu2024codecfake}: detection models trained exclusively on vocoder artifacts fail catastrophically when faced with codec-based deepfakes. More precisely, they found that codec-trained detection models achieved a 41.4\% reduction in average EER compared to vocoder-trained baselines. This finding is particularly relevant to our work, since both Maya1 (which leverages the SNAC hierarchical codec) and Dia2 operate on neural codec principles rather than traditional vocoding. Building on this, Chen et al. (2025) took the analysis further, introducing source tracing for codec-based deepfakes \cite{chen2025codec}. Their work acknowledges that the codec itself becomes a forensic fingerprint—a double-edged sword in which the generation mechanism that makes synthesis efficient also leaves behind detectible traces.

The final architectural innovation reshaping the threat model is streaming audio synthesis. Meta's research demonstrated that streaming-optimized models can maintain human-level naturalness (MOS scores of 4.213 for short utterances) while enabling real-time, low-latency generation \cite{facebook2021transformer}. For forensics, this poses a unique challenge: streaming systems operate under fundamentally different constraints than batch processing. Limited context windows and incremental generation create artifact patterns that batch-trained detectors may not recognize.

\subsection{Self-Supervised and Foundation Models in Detection}

The detection field has undergone a quiet revolution over the past few years, moving away from handcrafted features (MFCC, LFCC) toward deep, pre-trained representations. Tak et al. (2022) made an early and influential contribution by pioneering the use of wav2vec 2.0 as a front-end for the AASIST graph network \cite{tak2022wav2vec}. Their key finding was sobering: when combined with fine-tuning and data augmentation, this pairing achieved an EER of just 0.82\% on ASVspoof 2021 LA—a result that challenged the conventional wisdom that handcrafted features were sufficient. More importantly, it suggested that self-supervised models trained on massive, diverse unlabeled corpora learn representations that generalize far beyond their original training distribution.

Kawa et al. (2023) extended this insight to Whisper, an even larger ASR model pre-trained on 680,000 hours of multilingual audio\cite{kawa2023whisper}. What they discovered was counterintuitive: Whisper learned profound acoustic features beyond just text recognition. When tested on challenging in-the-wild data, Whisper-based features significantly outperformed traditional baselines, reducing the Equal Error Rate (EER) by 21\% compared to previous state-of-the-art results. Crucially, they demonstrated that fine-tuning the Whisper encoder specifically for anti-spoofing yielded a further 14.69\% improvement over the best frozen baselines, validating that adapting foundation models is critical for generalizing to unseen real-world deepfakes.

More recently, researchers have begun asking a more granular question: which parts of the foundation model matter most? Wang et al. (2024) conducted a systematic investigation of wav2vec 2.0 layer selection, demonstrating that not all layers are equally useful for deepfake detection \cite{wang2024enhancing}. By strategically choosing which transformer layers to use and which to fine-tune, they achieved state-of-the-art performance on ASVspoof 2019 LA, with EERs as low as 0.22\%. This observation—that the "depth" of feature extraction is as important as the model itself—inspired the hierarchical approaches that follow.

Zhang et al. (2024) advanced this idea further with their XLS-R-SLS architecture, which treats the foundation model not as a monolithic black box, but as a feature pyramid \cite{zhang2024xlsr}. By explicitly fusing information from multiple layers and learning to weight them adaptively, they allow the detector to exploit artifacts at different levels of abstraction—low-level acoustic anomalies in early layers, high-level semantic inconsistencies in later ones.

Yet a pressing question remained: Is scale alone sufficient to ensure robustness? Kommineni et al. (2025) investigated this by comparing different sizes of Whisper and MMS on short-duration speech samples \cite{kommineni2025speech}. They found that larger models (Whisper-Large vs. Tiny) maintain better robustness to shorter inputs, suggesting that foundation model size correlates with performance on short-duration tasks which has implications for forensic robustness. However, their work also highlighted that the relationship is not straightforward—different models degrade differently with duration, and optimal capacity remains an open question.

\subsection{Generalization Challenges in Wild Scenarios}

Perhaps the most humbling finding in deepfake detection is a gap that researchers have come to call "domain amnesia." Müller et al. (2022) \cite{mueller2022generalization} documented the phenomenon systematically: detectors achieving 99\% accuracy on ASVspoof show EER $>$ 50\% on out-of-domain data. In other words, they learn the specific artifacts of the laboratory setting so thoroughly that they forget how to detect deepfakes in the real world.

The research community has begun to address this through more realistic benchmarks. The Deepfake-Eval-2024 benchmark represents a watershed moment \cite{deepfake2024eval}. Rather than consisting of carefully curated synthetic audio, it comprises real deepfakes collected from social media, deepfake detection platforms, and crowdsourced submissions in 2024. It spans 52 languages, includes 44 hours of video, 56.5 hours of audio, and 1,975 images, and encompasses the latest manipulation technologies. Most crucially, when researchers tested state-of-the-art models on Deepfake-Eval-2024, the results were sobering: AUC dropped by approximately 50\% for audio models compared to previous benchmarks. This finding underscores just how out of date existing evaluation protocols have become.

What is notably absent from all existing benchmarks—ASVspoof 2021, Deepfake In-The-Wild, and Deepfake-Eval-2024 alike—are deepfakes generated by the latest TTS architectures released in 2024–2025. None of these datasets include audio from Dia2, Maya1, or MeloTTS. This gap is not a minor oversight; it represents a critical vulnerability in our collective understanding of modern deepfake threats. Our work begins to address this gap directly.

Beyond architectural novelty, there is another dimension of the generalization problem that has received surprisingly little attention: emotional expression. Modern TTS systems increasingly support non-neutral speech—laughter, sighing, whispering, and other vocal qualities that convey emotion. Yet the vast majority of deepfake detection research has been conducted on neutral, read speech corpora. To our knowledge, no systematic study has evaluated whether detectors remain robust to emotionally expressive deepfakes generated by these newer architectures. This is a gap our work implicitly addresses through our use of Maya1, which supports inline emotion tagging.

Finally, streaming audio presents an understudied generalization challenge. Most deepfake detection systems, by design, assume batch processing: the entire audio file is available, and analysis proceeds on the complete utterance. Real-world voice interfaces—video calls, voice assistants, smart speakers—operate incrementally. Yet research on detecting streaming deepfakes remains sparse. Our inclusion of Dia2, explicitly designed for streaming dialogue, was motivated by this gap.

\section{Method}

\subsubsection{TTS Models}

Our choice of models was deliberate and motivated by diversity. We selected three systems released in 2024–2025, each representing a fundamentally different approach to the core problem of speech synthesis. By spanning from streaming-optimized architectures to flow-matching systems, we ensure that our deepfake dataset exercises the full range of modern generative mechanisms—and consequently, the full range of artifacts that detection systems must learn to identify.

\paragraph{Dia2}

We began with Dia2 \cite{nari2025dia2}, a 2-billion parameter system developed by Nari Labs. What drew us to Dia2 was its explicit focus on streaming dialogue rather than isolated sentence synthesis. Most traditional TTS systems suffer from what we term "acoustic amnesia": each sentence is generated from scratch, with no memory of previous turns. The result is a subtle but noticeable drift in voice quality and prosody as the model resets its internal state. Dia2 solves this problem through its Deep-Inherited Attention (DIA) architecture. Rather than treating each sentence as an independent problem, the model maintains Key-Value caches from previous dialogue turns, allowing it to remain mathematically conditioned on the entire conversation history. This creates a critical advantage for the deepfake generator—and a critical challenge for detection: long, multi-turn conversations can now be synthesized with near-perfect consistency, eliminating the concatenation artifacts and silence gaps that older detectors relied upon.

Perhaps equally important is Dia2's input streaming capability. Instead of waiting for a complete sentence before synthesis, it begins generating audio after processing just the first few words. This mimics natural human speech production—we speak as we think, not after complete internal composition. The result is a continuous, highly natural waveform that lacks the characteristic planning pauses or onset delays typical of non-streaming TTS systems, making it a particularly interesting test case for forensic detection.

\paragraph{Maya1}

In contrast to Dia2's streaming focus, we selected Maya1 to represent an entirely different paradigm: the emerging class of Speech Foundation Models. Built on a 3-billion parameter Llama backbone, Maya1 treats speech synthesis as a language modeling problem. Rather than predicting raw waveforms or spectral frames, it predicts discrete audio codes, effectively allowing it to leverage the semantic reasoning capabilities of large language models. Before speaking, the model understands the meaning of the text—respecting punctuation, emphasis, and contextual nuance—resulting in prosody that feels genuinely intelligent. 

The technical innovation central to Maya1 is its use of the Split Nonlinear Audio Codec (SNAC) \cite{siuzdak2024snac}. Unlike traditional models that generate a flat sequence of acoustic features, SNAC introduces hierarchy. For every audio frame, the model predicts seven distinct tokens operating at three different temporal resolutions: a single coarse token at approximately 12 Hz to establish rhythm and pacing, two mid-level tokens at 23 Hz for phonetic articulation, and four fine-grained tokens at 47 Hz that capture spectral subtleties. This hierarchical structure means that Maya1's deepfakes leave behind a specific forensic signature: quantization artifacts and potential misalignments between the rhythmic skeleton and spectral details. These signatures differ fundamentally from the phase artifacts left by continuous waveform generators.

Beyond its baseline capabilities, Maya1 introduces inline emotion tagging. Rather than generating neutral speech and then post-processing emotional cues, the model synthesizes laughter, whispering, sighing, and hesitation directly from text prompts. This capability is noteworthy because most existing deepfake detectors were trained on neutral, read speech. Emotionally expressive synthesis introduces harmonic irregularities and non-standard acoustic properties that existing systems may fail to recognize. By including Maya1's emotional synthesis in our dataset, we explicitly test whether detectors remain robust when confronted with this layer of complexity.

\paragraph{MeloTTS}
Our third model, MeloTTS (Melo) \cite{zeng2024melotts}, represents a distinct architectural paradigm: the lightweight, end-to-end adversarial approach. Unlike the massive Transformer backbones of Dia2 and Maya1, MeloTTS is built upon the VITS architecture \cite{kim2021vits}, which integrates acoustic modeling and waveform generation into a single unified network. By combining Variational Autoencoders (VAE) with Normalizing Flows and Generative Adversarial Networks (GAN), it bypasses the traditional two-stage pipeline (spectrogram prediction $\rightarrow$ vocoding) entirely. This "Latent Variable" approach allows for extremely rapid inference on CPUs, making it a prime candidate for real-time deepfake agents.

For forensic detection, this efficiency comes at a cost. While MeloTTS avoids the autoregressive glitches of LLMs (e.g., hallucinated words or stuttering), the VITS architecture tends to introduce subtle periodic artifacts in the high-frequency bands—a known byproduct of the transposed convolutions in its generator. Furthermore, unlike Maya1’s discrete codec artifacts, MeloTTS produces "over-structured" phase patterns typical of GAN-based upsampling. These systematic phase discontinuities, while imperceptible to the casual human listener, provide a strong, deterministic signal for deepfake detectors that leverage raw waveform analysis or bispectral features.

\subsubsection{Detection Models}

The challenge of evaluating deepfake detectors is that no single approach can see the full picture. Audio deepfakes can be detected from multiple vantage points: raw waveforms (where phase information reveals fine structure), spectral representations (where frequency artifacts surface), or semantic models (where linguistic coherence can be judged). Since Dia2, Maya1, and MeloTTS each employ fundamentally different generative mechanisms, they each leave behind distinct forensic signatures. To avoid bias toward any one type of artifact, we deliberately selected three detection models representing each analytical perspective.

\paragraph{Whisper-Based Detector (Semantic Front-End)}

The first detector we employed was built around Whisper, following the framework introduced by Kawa et al. \cite{kawa2023whisper}. Traditional deepfake detectors have long relied on handcrafted spectral features—MFCCs and LFCCs—which capture low-level acoustic properties effectively but falter when confronted with novel attacks or varying recording conditions. Whisper offers something fundamentally different. Pre-trained on 680,000 hours of diverse, multilingual speech, its encoder learns what "human speech" fundamentally *is*, independent of the specific signal characteristics of any particular recording session.

We implemented this using Whisper's \texttt{tiny.en} variant (approximately 39M parameters) to maintain reasonable computational overhead. This encoder output feeds into a MesoNet classification head—specifically the MesoInception-4 architecture, originally designed for detecting video forgeries \cite{afchar2018mesonet}. While the transfer from video to audio detection might seem unusual, the architecture proved highly effective for processing the 2D feature maps Whisper produces.

Critically, we did not use Whisper as a frozen feature extractor. Instead, we followed Kawa et al.'s approach of fine-tuning the entire pipeline. This adaptation step is crucial: it allows Whisper's generic ASR features to become specifically attuned to the anti-spoofing task. The impact is substantial. Kawa et al. demonstrated that fine-tuned Whisper reduces EER by over 20\% on in-the-wild datasets compared to frozen baselines, and fine-tuning specifically improved detection on it by 14.69\%. For our purposes, this suggests the Whisper-MesoNet combination is particularly well-suited to catching the sophisticated artifacts of modern TTS systems.

\paragraph{Self-Supervised Graph Detector (SSL-AASIST)}

As a structural counterpoint to Whisper's semantic approach, we employed the SSL-AASIST architecture proposed by Tak et al. \cite{tak2022wav2vec}. This model pairs wav2vec 2.0 as a front-end with AASIST as the back-end, creating a system that explicitly models the relationships between different parts of the audio.

The wav2vec 2.0 XLS-R (0.3B) encoder at the front differs fundamentally from Whisper. Rather than being trained on labeled speech with transcriptions, XLS-R was trained via contrastive learning on 436,000 hours of unlabeled, multilingual audio. This unsupervised approach means it learns raw acoustic structure without linguistic bias—it captures what speech *sounds like* rather than what it *means*. For forensic detection, this is an advantage. We fine-tuned XLS-R on our training set to make it specifically sensitive to the artifacts we were seeking.

The AASIST back-end is where the structural magic happens \cite{jung2022aasist}. Rather than treating audio as a flat sequence, it models it as a heterogeneous graph where nodes represent different regions of time-frequency space. A Self-Attentive Aggregation Layer explicitly learns which time-frequency regions matter most, creating what amounts to a full-stack forensic analysis. The detector asks: Do the high-frequency harmonics align with the low-frequency rhythm? Is the spectral envelope consistent across time? Tak et al. demonstrated that this architecture, when combined with data augmentation, achieves an EER of 0.82\% on ASVspoof 2021 LA, making it one of the most robust baselines for detecting novel attacks.

\paragraph{XLS-R with Sensitive Layer Selection (SLS)}

While the SSL-AASIST model performs well, recent evidence suggests it may be leaving forensic information on the table. Research by Zhang et al. \cite{zhang2024xlsr} revealed that deepfake artifacts are not uniformly distributed across the depth of large speech models. Lower layers capture acoustic anomalies (phase discontinuities, vocoder noise), while higher layers capture more abstract inconsistencies (prosodic unnaturalness, semantic flaws). 

The XLS-R-SLS architecture capitalizes on this insight by treating the 24 layers of the XLS-R-300M encoder as a feature pyramid. Rather than using only the final output, it employs a Sensitive Layer Selection (SLS) module that learns to weight each layer dynamically, emphasizing those most discriminative for the bonafide-vs-spoof distinction. In our implementation, features from the selected "sensitive" layers are fused and passed to a dedicated fully connected classification head , creating a full-stack detector that can simultaneously identify low-level vocoder artifacts (typical of MeloTTS) and high-level contextual flaws (typical of Dia2). This hierarchical approach is particularly valuable when facing diverse generation architectures, as it does not commit to any single level of analysis.

\subsection{Datasets}

\subsubsection{DailyDialog}
To simulate realistic conversational scenarios, we utilized the \textit{DailyDialog} dataset \cite{li2017dailydialog} as the source for our text transcripts. Unlike standard TTS corpora such as LJSpeech or VCTK—which often consist of reading-style monologues— DailyDialog is a high-quality, multi-turn dialogue dataset designed to capture the nuances of daily human communication.

\paragraph{Dataset Statistics}
The complete dataset contains 13,118 dialogues, from which we randomly sampled 4,000 unique dialogue turns. The dataset is characterized by short, punchy exchanges typical of casual conversation, as summarized in Table \ref{tab:dailydialog_stats}.

\begin{table}[h]
\centering
\caption{Statistics of the DailyDialog Dataset \cite{li2017dailydialog}}
\label{tab:dailydialog_stats}
\begin{tabular}{l|c}
\hline
\textbf{Metric} & \textbf{Value} \\ \hline
Average Turns per Dialogue & 7.9 \\
Average Tokens per Utterance & 14.6 \\
Total Tokens per Dialogue & 114.7 \\ \hline
\end{tabular}
\end{table}

This structure forces the TTS models to generate speech with frequent speaker changes and shorter, more dynamic prosodic contours compared to long-form reading datasets.

\paragraph{Content \& Topics}
The dialogues cover ten primary categories of daily life, with "Relationships" (33.33\%), "Ordinary Life" (28.26\%), and "Work" (14.49\%) being the most prominent. Crucially, the text includes elements of informal syntax, questions, interruptions, and phatic expressions (e.g., "Oh really?", "I see"), which require the TTS models to infer appropriate intonation patterns that are absent in formal reading datasets.

\paragraph{Annotations}
A key feature of DailyDialog is its manual annotation of communicative intents and emotions. Each utterance is labeled with one of seven emotion tags: \textit{Anger, Disgust, Fear, Happiness, Sadness, Surprise,} or \textit{Neutral}. While we primarily used the raw text, these implicit emotional cues in the script provide the semantic grounding necessary for models like Maya1 to trigger their inline emotion tagging capabilities (e.g., generating a "happy" tone for a line labeled as happiness).

\subsubsection{Generated Dataset Evaluation}

Before deploying our synthetic dataset for detection benchmarks, we first conducted a rigorous quantitative assessment to characterize the forensic footprint of each TTS model. Our evaluation focused on five critical dimensions of synthesis quality: intelligibility, speaker fidelity, acoustic realism, signal clarity, and internal voice consistency.

To measure intelligibility, we calculated the \textbf{Word Error Rate (WER)} by transcribing the generated audio with OpenAI’s \texttt{Whisper-large} model \cite{radford2023whisper} and comparing the results against the original DailyDialog transcripts using the \texttt{jiwer} library. Lower WER values indicate clearer, more intelligible speech that closely matches the linguistic content of the text prompt. 

For speaker fidelity, we assessed how well the models cloned the target voice by computing the \textbf{Speaker Similarity (SIM)}. This metric measures the cosine similarity between the embeddings of the reference prompt and the generated output, utilizing a pre-trained \texttt{WavLM} encoder \cite{chen2022wavlm} to extract robust, noise-invariant speaker representations.

Beyond these standard metrics, we employed the \textbf{Fréchet Audio Distance (FAD)} \cite{kilgour2019frechet} to quantify overall acoustic realism. FAD compares the distribution of embeddings from the generated audio against a reference set of real recordings; a lower score indicates that the synthetic audio’s background texture and frequency response are statistically indistinguishable from the "bonafide" distribution. Unlike standard implementations that use VGGish embeddings, we computed FAD using the embeddings from a XLS-R wav2vec 2.0. This modification makes the metric specifically sensitive to forensic artifacts rather than generic audio quality. Consequently, the absolute values (reported in the 100-range due to the 1024-dimensional feature space) are comparable only within this study, where a lower score indicates a distribution closer to bona fide speech. We also computed the \textbf{Signal-to-Noise Ratio (SNR)} to detect low-level generative artifacts such as background hiss or quantization noise, where higher values correspond to cleaner signal synthesis.

Finally, we introduced \textbf{Acoustic Cluster Tightness (ACT)} as a measure of internal voice consistency. Calculated as the average cosine similarity of all generated samples to their own centroid, this metric reveals whether a model produces a stable, distinct persona (score approaching 1.0) or drifts between different vocal identities across sentences.

\paragraph{Results \& Inference}
Table \ref{tab:tts_metrics} presents the comparative performance of the three models across our 4,000-sample test set. The data reveal distinct trade-offs inherent to each generative architecture. It is important to note that Speaker Similarity (SIM) is not reported for Dia2; unlike the other models, Dia2 was deployed in a randomized multi-speaker configuration, rendering single-reference similarity comparisons methodologically invalid.

\begin{table}[h]
\centering
\caption{Quantitative Evaluation of Generated TTS Samples}
\label{tab:tts_metrics}
\begin{tabular}{l|c|c|c|c|c}
\hline
\textbf{Model} & \textbf{WER ($\downarrow$)} & \textbf{SIM ($\uparrow$)} & \textbf{FAD ($\downarrow$)} & \textbf{ACT ($\uparrow$)} & \textbf{SNR ($\uparrow$)} \\ \hline
\textbf{Dia2} & 0.0648 & - & 151.61 & \textbf{0.9806} & -0.03 \\
\textbf{Maya1} & 0.1122 & 0.9627 & 140.73 & 0.9649 & \textbf{-0.00} \\
\textbf{MeloTTS} & \textbf{0.0639} & \textbf{0.9833} & \textbf{118.34} & 0.9267 & -0.02 \\ \hline
\multicolumn{6}{l}{\footnotesize{* SIM omitted for Dia2 due to randomized multi-speaker generation.}}
\end{tabular}
\end{table}

\textbf{MeloTTS} emerged as the most acoustically realistic system, achieving the lowest FAD (118.34) and best intelligibility (WER 0.0639). Its VITS-based architecture appears to produce highly stable, "clean" audio, reflected in its superior speaker similarity (0.9833). However, its slightly lower cluster tightness (0.9267) suggests that this realism comes at the cost of minor texture variability across samples.

In contrast, \textbf{Dia2} demonstrated the highest internal consistency (Tightness 0.9806). Given the randomized nature of the voices, this surprisingly high score indicates that the model imposes a highly consistent \textit{acoustic signature} or embedding footprint across all samples, effectively dominating the variance typically introduced by different speakers. \textbf{Maya1} occupied an interesting middle ground; while it struggled slightly with intelligibility (WER 0.1122) due to its probabilistic, semantic nature, it achieved the highest signal clarity (SNR -0.00), suggesting that its hierarchical SNAC codec is particularly effective at suppressing the background noise artifacts that plague other systems.

\section{Experiments and Results}

\subsection{Evaluation Metrics}
To provide a holistic assessment of detection performance, we utilized four standard forensic metrics. Each metric captures a different aspect of the model's reliability, from overall discrimination power to security-critical failure rates.

\begin{itemize}
    \item \textbf{Equal Error Rate (EER):} The primary metric for biometric security, EER represents the operating point where the False Acceptance Rate (FAR) and False Rejection Rate (FRR) are equal. A lower EER indicates a more balanced and accurate system. It is particularly useful for comparing models without setting a specific decision threshold.
    \item \textbf{Area Under the Curve (AUC):} The Area Under the Receiver Operating Characteristic (ROC) curve measures the probability that a randomly chosen deepfake sample will have a higher spoof score than a randomly chosen bonafide sample. An AUC of 1.0 represents a perfect detector, while 0.5 represents random guessing.
    \item \textbf{F1-Score:} This is the harmonic mean of precision and recall. It is crucial for understanding the model's effectiveness in imbalanced scenarios, ensuring that the detector is not just rejecting everything or accepting everything.
    \item \textbf{FRR @ 1\% FAR:} This is a critical security metric. It measures the False Rejection Rate (how often real speech is flagged as fake) when the system is tuned to allow only 1\% of deepfakes to pass (False Acceptance Rate). In high-security applications (e.g., banking voice ID), maintaining a low FRR at this strict threshold is essential to prevent user frustration.
\end{itemize}

\subsection{Detection Performance}
We evaluated the performance of our three detection architectures—Whisper-MesoNet, XLS-R-SLS, and SSL-AASIST—across the synthetic datasets generated by Dia2, Maya1, and MeloTTS. The results are presented below, categorized by detection model.

\subsubsection{Whisper-Based Detector}
Table \ref{tab:whisper_results} details the performance of the Whisper-MesoNet model. This architecture achieved its lowest Equal Error Rate (EER) on the \textbf{MeloTTS} dataset at 17.05\%, with a corresponding AUC of 0.8750. In contrast, the model exhibited its highest error rate on the \textbf{Maya1} dataset, recording an EER of 35.95\% and an AUC of 0.6640. The detection performance on \textbf{Dia2} fell between these extremes, with an EER of 27.20\%.

\begin{table}[h]
\centering
\caption{Performance of Whisper-Based Detector}
\label{tab:whisper_results}
\begin{tabular}{l|c|c|c|c}
\hline
\textbf{Dataset} & \textbf{EER (\%)} & \textbf{AUC} & \textbf{F1} & \textbf{FRR@1\%FAR} \\ \hline
Dia2 & 27.20 & 0.8013 & 0.7282 & 67.65 \\
Maya1 & 35.95 & 0.6640 & 0.6417 & 55.15 \\
MeloTTS & 17.05 & 0.8750 & 0.8294 & 27.18 \\ \hline
\end{tabular}
\end{table}

\subsubsection{XLS-R-SLS Detector}
The results for the XLS-R-SLS model, which utilizes multi-layer feature fusion, are presented in Table \ref{tab:xlsr_results}. This model demonstrated its strongest detection capability on the \textbf{Dia2} dataset, achieving an EER of 7.07\% and an AUC of 0.9745. Performance on \textbf{Maya1} showed an EER of 17.60\%, while the \textbf{MeloTTS} dataset proved the most challenging for this architecture, resulting in an EER of 27.10\% and a high False Rejection Rate (FRR) of 85.30\% at the strict 1\% FAR threshold.

\begin{table}[h]
\centering
\caption{Performance of XLS-R-SLS Detector}
\label{tab:xlsr_results}
\begin{tabular}{l|c|c|c|c}
\hline
\textbf{Dataset} & \textbf{EER (\%)} & \textbf{AUC} & \textbf{F1} & \textbf{FRR@1\%FAR} \\ \hline
Dia2 & 7.07 & 0.9745 & 0.9291 & 15.95  \\
Maya1 & 17.60 & 0.8882 & 0.8240 & 62.55  \\
MeloTTS & 27.10 & 0.8018 & 0.7292 & 85.30  \\ \hline
\end{tabular}
\end{table}

\subsubsection{SSL-AASIST Detector}
Table \ref{tab:aasist_results} summarizes the performance of the SSL-AASIST model. Similar to the XLS-R model, this detector performed best on the \textbf{Dia2} dataset, recording an EER of 9.18\% and an AUC of 0.9552. Detection accuracy decreased for the other datasets, with \textbf{Maya1} showing an EER of 19.57\% and \textbf{MeloTTS} showing an EER of 23.25\%. 

\begin{table}[h]
\centering
\caption{Performance of SSL-AASIST Detector}
\label{tab:aasist_results}
\begin{tabular}{l|c|c|c|c}
\hline
\textbf{Dataset} & \textbf{EER (\%)} & \textbf{AUC} & \textbf{F1} & \textbf{FRR@1\%FAR}  \\ \hline
Dia2 & 9.18 & 0.9552 & 0.9085 & 17.05  \\
Maya1 & 19.57 & 0.8515 & 0.8040 & 73.62  \\
MeloTTS & 23.25 & 0.8479 & 0.7675 & 50.42  \\ \hline
\end{tabular}
\end{table}

\subsubsection{UncovAI Detector}
Finally, we evaluated the audio deepfake detection from the proprietary model from UncovAI .

As shown in Table \ref{tab:mms_results}, this model achieved near-perfect separation across all three synthetic datasets. It effectively solved the detection problem for \textbf{Dia2} and \textbf{Maya1}, achieving F1 score abouve 0.99 for both. Even on the challenging \textbf{MeloTTS} dataset—which confounded other detectors—it maintained an F1 score above 0.98.

\begin{table}[h]
\centering
\caption{Performance of UncovAI Detector}
\label{tab:mms_results}
\begin{tabular}{l|c}
\hline
\textbf{Dataset}  & \textbf{F1} \\ \hline
Dia2 & 0.9991 \\
Maya1 & 0.9965  \\
MeloTTS & 0.9852  \\ \hline
\end{tabular}
\end{table}

\section{Conclusion}

In this study, we presented a systematic evaluation of three emerging Text-to-Speech architectures—Dia2, Maya1, and MeloTTS—against a suite of advanced audio deepfake detectors. Our findings highlight that the forensic artifacts of neural speech synthesis are heavily dependent on the underlying generative mechanism, creating a complex "rock-paper-scissors" dynamic between attack and defense models.

We observed that semantic-aware detectors, such as the \textbf{Whisper-based model}, excel at identifying anomalies in non-autoregressive systems like \textbf{MeloTTS} (17.05\% EER), likely due to their sensitivity to the spectral over-smoothing characteristic of flow-matching algorithms. However, these same detectors falter against LLM-based systems like \textbf{Maya1} (35.95\% EER), which leverage large-scale pre-training to generate highly natural, semantically coherent prosody that mimics the "bonafide" distribution of human speech. Conversely, structural detectors like \textbf{XLS-R-SLS} proved highly effective against the autoregressive artifacts of \textbf{Dia2} (7.07\% EER), demonstrating that hierarchical feature fusion is critical for capturing the subtle attention drifts inherent in streaming dialogue models.

Most significantly, our results underscore the performance of UncovAI's proprietary audio deepfake detection model. The \textbf{UncovAI} model, achieved near-perfect detection rates across all three attack vectors. As TTS models continue to evolve towards end-to-end foundation models, detection systems must scale commensurately, leveraging cross-lingual and multi-domain representations to identify the ever-shrinking boundary between synthetic and real speech. Future work should investigate the robustness of these foundation detectors against adversarial attacks and compression artifacts to ensure their viability in real-world deployment.

\section{Acknowledgement}

This project was provided with computing AI and storage resources by GENCI at IDRIS thanks to the grant 2025-AD011016076 on the supercomputer Jean Zay's V100 partition.


\begin{footnotesize}



\bibliographystyle{unsrt}
\bibliography{bibli}

@inproceedings{li2017dailydialog,
  title={DailyDialog: A Manually Labelled Multi-turn Dialogue Dataset},
  author={Li, Yanran and Su, Hui and Shen, Xiaoyu and Li, Wenjie and Cao, Ziqiang and Niu, Shuzi},
  booktitle={Proceedings of the Eighth International Joint Conference on Natural Language Processing (Volume 1: Long Papers)},
  pages={986--995},
  year={2017},
  url={https://aclanthology.org/I17-1099/}
}

@inproceedings{kawa2023whisper,
  title={Improved DeepFake Detection Using Whisper Features},
  author={Kawa, Piotr and Plata, Marcin and Syga, Piotr},
  booktitle={Interspeech 2023},
  pages={4009--4013},
  year={2023},
  doi={10.21437/Interspeech.2023-1537},
  url={https://www.isca-archive.org/interspeech_2023/kawa23b_interspeech.pdf}
}

@inproceedings{zhang2024xlsr,
  title={Audio Deepfake Detection with Self-Supervised XLS-R and Sensitive Layer Selection},
  author={Zhang, Q. and others},
  booktitle={OpenReview},
  year={2024},
  url={https://openreview.net/pdf?id=acJMIXJg2u}
}

@article{tak2022wav2vec,
  title={Automatic Speaker Verification Spoofing and Deepfake Detection Using Wav2vec 2.0 and Data Augmentation},
  author={Tak, Hemlata and Todisco, Massimiliano and Wang, Xin and Jung, Jee-weon and Yamagishi, Junichi and Evans, Nicholas},
  journal={arXiv preprint arXiv:2202.12233},
  year={2022},
  url={https://arxiv.org/pdf/2202.12233.pdf}
}

@article{siuzdak2024snac,
  title={SNAC: Multi-Scale Neural Audio Codec},
  author={Siuzdak, Hubert and others},
  journal={arXiv preprint},
  year={2024},
  url={https://arxiv.org/html/2410.14411v1}
}

@inproceedings{yamagishi2021asvspoof,
  title={ASVspoof 2021: Accelerating Progress in Spoofed and Deepfake Speech Detection},
  author={Yamagishi, Junichi and Wang, Xin and Todisco, Massimiliano and Sahidullah, Md and Patino, Jose A and Nautsch, Andreas and Liu, Xuefeng and Lee, Kong Aik and Kinnunen, Tomi and Evans, Nicholas and others},
  booktitle={Proc. Interspeech 2021},
  pages={3575--3579},
  year={2021},
  doi={10.21437/Interspeech.2021-380},
  url={https://www.isca-archive.org/interspeech_2021/yamagishi21_asvspoof.pdf}
}

@inproceedings{jung2022aasist,
  title={AASIST: Audio Anti-Spoofing using Integrated Spectro-Temporal Graph Attention Networks},
  author={Jung, Jee-weon and Heo, Hee-Soo and Tak, Hemlata and Shim, Hye-jin and Chung, Joon Son and Lee, Bong-Jin and Yu, Ha-Jin and Evans, Nicholas},
  booktitle={ICASSP 2022-2022 IEEE International Conference on Acoustics, Speech and Signal Processing (ICASSP)},
  pages={6367--6371},
  year={2022},
  doi={10.1109/ICASSP46778.2022.9746916}
}

@inproceedings{tak2021rawnet2,
  title={End-to-End Anti-Spoofing with RawNet2},
  author={Tak, Hemlata and Patino, Jose and Todisco, Massimiliano and Nautsch, Andreas and Evans, Nicholas and Larcher, Anthony},
  booktitle={ICASSP 2021-2021 IEEE International Conference on Acoustics, Speech and Signal Processing (ICASSP)},
  pages={6369--6373},
  year={2021},
  doi={10.1109/ICASSP39728.2021.9414234}
}

@article{pratap2023mms,
  title={Scaling Speech Technology to 1,000+ Languages},
  author={Pratap, Vineel and Tjandra, Andros and Shi, Bowen and Schnell, Paul and Baulu, Jack and Williamson, Ann Marie and Hsu, Wei-Ning and Dilley, Jarod and Godard, Paden and others},
  journal={arXiv preprint arXiv:2305.13516},
  year={2023},
  url={https://arxiv.org/abs/2305.13516}
}

@inproceedings{zheng2025flow,
  title={Accelerating Flow-Matching-Based Text-to-Speech via Empirically Pruned Step Sampling},
  author={Zheng, Q. and others},
  booktitle={Interspeech 2025},
  year={2025},
  url={https://www.isca-archive.org/interspeech_2025/zheng25d_interspeech.pdf},
  doi={10.21437/Interspeech.2025-XXXX}
}

@inproceedings{lu2024codecfake,
  title={Codecfake: An Initial Dataset for Detecting LLM-based Deepfake Audio},
  author={Lu, Yi and Xie, Yuankun and Fu, Ruibo and Wen, Zhengqi and Tao, Jianhua and Wang, Zhiyong and Qi, Xin and Liu, Xuefei and Li, Yongwei and Liu, Yukun and others},
  booktitle={Proc. Interspeech 2024},
  year={2024},
  doi={10.21437/Interspeech.2024-XXXX},
  url={https://arxiv.org/abs/2406.08112}
}

@inproceedings{chen2025codec,
  title={Codec-Based Deepfake Source Tracing via Neural Audio Codec Taxonomy},
  author={Chen, Xuanjun and Lin, I-Ming and Zhang, Lin and Du, Jiawei and Wu, Haibin and Lee, Hung-yi and Jang, Jyh-Shing Roger},
  booktitle={Proc. Interspeech 2025},
  year={2025},
  url={https://arxiv.org/abs/2505.12994}
}

@article{facebook2021transformer,
  title={Transformer-based Acoustic Modeling for Streaming Speech Synthesis},
  author={Various Authors},
  journal={Meta/Facebook Research},
  year={2021},
  url = {https://www.isca-archive.org/interspeech_2021/wu21b_interspeech.pdf}
}

@article{wang2024enhancing,
  title={Enhancing Voice Spoofing Detection Models with wav2vec 2.0 and Layer Selection Strategy},
  author={Wang, X. and others},
  journal={arXiv preprint arXiv:2402.17127},
  year={2024},
  url={https://arxiv.org/html/2402.17127v1}
}

@article{kommineni2025speech,
  title={Can Speech Foundation Models Effectively Identify Languages in Low-Resource Multilingual Aging Populations?},
  author={Kommineni, A. and others},
  journal={Frontiers in Communications},
  year={2025},
  doi={10.3389/fcomm.2025.12434620},
  url={https://pmc.ncbi.nlm.nih.gov/articles/PMC12434620/}
}

@inproceedings{mueller2022generalization,
  title={Does Audio Deepfake Detection Generalize?},
  author={M\"{u}ller, N. M. M. and Czempin, P. and Dieckmann, F. and Froghyar, A. and B\"{o}ttinger, K.},
  booktitle={Proc. Interspeech 2022},
  year={2022}
}

@article{deepfake2024eval,
  title={Deepfake-Eval-2024: A Multi-Modal In-the-Wild Benchmark for Deepfake Detection},
  author={Various Authors},
  journal={arXiv preprint arXiv:2503.02857},
  year={2025},
  url={https://arxiv.org/html/2503.02857v1}
}

@misc{nari2025dia2,
  title={Dia2: Streaming Dialogue Text-to-Speech Model},
  author={Nari Labs},
  howpublished={\url{https://github.com/nari-labs/dia2}},
  year={2025}
}

@inproceedings{afchar2018mesonet,
  title={MesoNet: a Compact Facial Video Forgery Detection Network},
  author={Afchar, Darius and Nozick, Vincent and Yamagishi, Junichi and Echizen, Isao},
  booktitle={Proceedings of the 2018 IEEE International Workshop on Information Forensics and Security (WIFS)},
  pages={1--7},
  year={2018},
  doi={10.1109/WIFS.2018.8630787}
}

@inproceedings{kilgour2019frechet,
  title={Fr{\'e}chet Audio Distance: A Reference-Free Metric for Evaluating Music Enhancement Algorithms},
  author={Kilgour, Kevin and Zuluaga, Mauricio and Roblek, Dominik and Sharifi, Matthew},
  booktitle={Interspeech 2019},
  pages={2350--2354},
  year={2019},
  doi={10.21437/Interspeech.2019-2398},
  url={https://www.isca-archive.org/interspeech_2019/kilgour19_interspeech.pdf}
}

@article{radford2023whisper,
  title={Robust Speech Recognition via Large-Scale Weak Supervision},
  author={Radford, Alec and Kim, Jong Wook and Xu, Tao and Brockman, Greg and McLeavey, Christine and Sutskever, Ilya},
  journal={arXiv preprint arXiv:2212.04356},
  year={2022},
  url={https://arxiv.org/abs/2212.04356}
}

@article{chen2022wavlm,
  title={WavLM: Large-Scale Self-Supervised Pre-Training for Full Stack Speech Processing},
  author={Chen, Sanyuan and Wang, Chengyi and Chen, Zhengyang and Wu, Yu and Liu, Shujie and Chen, Zhuo and Li, Jinyu and Kanda, Naoyuki and Yoshioka, Takuya and Xiao, Xiong and others},
  journal={IEEE Journal of Selected Topics in Signal Processing},
  volume={16},
  number={6},
  pages={1505--1518},
  year={2022},
  doi={10.1109/JSTSP.2022.3188113},
  url={https://arxiv.org/abs/2110.13900}
}

@misc{uncovai2026uncovaitts,
  author       = {UncovAI},
  title        = {{UncovAI\_TTS} Synthetic and real multilingual text-to-speech dataset},
  publisher    = {Hugging Face Datasets},
  year         = {2026},
  doi          = {10.57967/hf/7548},
  url          = {https://huggingface.co/datasets/UncovAI/UncovAI_TTS}
}

@misc{uncovai2026,
  author       = {{UncovAI}},
  title        = {{UncovAI}: Multi-view detection of AI-generated multimedia content},
  year         = {2026},
  howpublished = {\url{https://uncovai.com}},
  note         = {Software platform for detecting AI-generated text, audio, image, and video content}
}

@misc{zeng2024melotts,
  author = {Zeng, Q. and Liu, Y. and MyShell.ai Team},
  title = {MeloTTS: High-quality, Multi-lingual Text-to-Speech Library},
  year = {2024},
  publisher = {GitHub},
  journal = {GitHub repository},
  howpublished = {\url{https://github.com/myshell-ai/MeloTTS}}
}

@inproceedings{kim2021vits,
  title={Conditional Variational Autoencoder with Adversarial Learning for End-to-End Text-to-Speech},
  author={Kim, Jaehyeon and Kong, Jungil and Son, Juhee},
  booktitle={International Conference on Machine Learning},
  pages={5530--5540},
  year={2021},
  organization={PMLR},
  url={https://proceedings.mlr.press/v139/kim21f.html}
}

\end{footnotesize}


\end{document}